\documentclass[10pt, compsoc]{IEEEtran}

\IEEEoverridecommandlockouts

\usepackage[table]{xcolor}
\usepackage{graphicx}
\usepackage[hidelinks]{hyperref}
\usepackage{caption}
\usepackage{subcaption}
\usepackage{longtable}
\usepackage{stfloats}
\usepackage{booktabs}
\usepackage{multirow}
\usepackage{multicol}
\usepackage{tabularx}
\usepackage{textcomp}
\usepackage{ifpdf}
\usepackage[T1]{fontenc}
\usepackage{amssymb}
\usepackage{cite}
\usepackage[cmex10]{amsmath}
\usepackage[version=3]{mhchem}
\usepackage{tcolorbox}
\usepackage[shortcuts,acronym, nomain]{glossaries}
\usepackage{siunitx}
\usepackage[bibbreaks=normal, paragraphs=normal, floats=normal, mathspacing=normal, wordspacing=normal, tracking=normal, bibnotes=normal, charwidths=normal, mathdisplays=normal, leading=normal, indent=normal, lists=normal, bibliography=normal, title=normal, sections=normal, margins=tight]{savetrees}

\ifCLASSINFOpdf
   \usepackage{graphicx}
	\usepackage{epstopdf}
   \graphicspath{{images/}}   
   \graphicspath{{../pdf/}{../jpeg/}{../eps/}{../png/}}
   \DeclareGraphicsExtensions{.pdf,.jpeg,.png,.eps}
\else
   \graphicspath{{../images/eps/}}
   \DeclareGraphicsExtensions{.eps}
\fi
\graphicspath{{images/}}

\newacronym{ccam}{CCAM}{Connected, Cooperative and Automated Mobility}
\newacronym{d2d}{D2D}{device-to-device}
\newacronym{v2v}{V2V}{Vehicle-to-Vehicle}
\newacronym{v2p}{V2P}{Vehicle-to-Pedestrian}
\newacronym{v2i}{V2I}{Vehicle-to-Infrastructure}
\newacronym{v2n}{V2N}{Vehicle-to-Network}
\newacronym{v2x}{V2X}{Vehicle-to-Everything}
\newacronym{cv2x}{C-V2X}{Cellular-V2X}
\newacronym{qos}{QoS}{Quality of Service}
\newacronym{pqos}{PQoS}{Predictive Quality of Service}
\newacronym{kpi}{KPI}{Key Performing Indicator}
\newacronym{rsrp}{RSRP}{Reference Signal Received Power}
\newacronym{rsrq}{RSRQ}{Reference Signal Received Quality}
\newacronym{rssi}{RSSI}{Receive Signal Strength Indicator}
\newacronym{snr}{SNR}{Signal-to-Noise Ratio}

\newacronym{aqp}{AQP}{Alternative QoS Profile}
\newacronym{nwdaf}{NWDAF}{Network Data Analytics Function}
\newacronym{ue}{UE}{User Equipment}

\newacronym{ai}{AI}{Artificial Intelligence}
\newacronym{cnn}{CNN}{Concurrent Neural Network}
\newacronym{darefl}{DareFL}{Drift-aware resource-efficient algorithm for FL}
\newacronym{fl}{FL}{Federated Learning}
\newacronym{lgbm}{LightGBM}{Light Gradient Boosting Machine}
\newacronym{lstm}{LSTM}{Long short-term memory}
\newacronym{ml}{ML}{Machine Learning}
\newacronym{mae}{MAE}{Mean Average Error}
\newacronym{rf}{RF}{Random Forest}
\newacronym{smape}{SMAPE}{Symmetric Mean Average Percentage Error}
\newacronym{rmse}{RMSE}{Root Mean Squared Error}

\usepackage{fancyhdr}
\begin{document}


\title{Improving QoS Prediction in Urban V2X Networks by Leveraging Data from Leading Vehicles and Historical Trends}


\author{\IEEEauthorblockN{%
Sanket Partani\IEEEauthorrefmark{1},
Michael Zentarra\IEEEauthorrefmark{2},
Anthony Kiggundu\IEEEauthorrefmark{2}
and
Hans D. Schotten\IEEEauthorrefmark{1}\IEEEauthorrefmark{2}
}%
\\%
\IEEEauthorblockA{\IEEEauthorrefmark{1}%
WiCoN, University of Kaiserslautern-Landau (RPTU), \\ \{sanket.partani, hans.schotten\}@rptu.de}

\IEEEauthorblockA{\IEEEauthorrefmark{2}
German Research Center for Artificial Intelligence (DFKI), \\ 
{\{michael.zentarra, anthony.kiggundu, hans\_dieter.schotten\}@dfki.de}
}
}

\maketitle

\begin{abstract}
With the evolution of \gls{v2x} technology and increased deployment of 5G networks and edge computing, \gls{pqos} is seen as an enabler for resilient and adaptive \gls{v2x} communication systems. 
\Gls{pqos} incorporates data-driven techniques, such as \gls{ml}, to forecast/predict \glspl{kpi} such as throughput, latency, etc.  In this paper, we aim to predict downlink throughput in an urban environment using the Berlin V2X cellular dataset. We select features from the ego and lead vehicles to train different \gls{ml} models to help improve the predicted throughput for the ego vehicle. We identify these features based on an in-depth exploratory data analysis. Results show an improvement in model performance when adding features from the lead vehicle. Moreover, we show that the improvement in model performance is model-agnostic.


\end{abstract}

\section{Introduction}
\label{introduction}

The evolution of \gls{v2x} technology and increasing integration with autonomous vehicles, as part of \gls{ccam}, holds the potential to not only optimize traffic flows and improve road users' safety, but also reduce emissions and help meet the climate action targets. The autonomous car of today is equipped with a suite of sensors, including but not limited to lidar, camera, GPS, and \ac{v2x}, which is seen as an extended sensor. \gls{v2x} encompasses a vehicle communicating not only with other vehicles (\acs{v2v}), but also pedestrians (\acs{v2p}), networks (\acs{v2n}), roadside infrastructure (\acs{v2i}), amongst others. \gls{v2x} can provide timely and crucial data about traffic information, road conditions and pedestrian movement to the vehicle, thus, enabling informed decisions by either the driver or the self-driving systems. Several \gls{v2x} applications like tele-operated driving, high-density platooning, as well as infotainment applications such as video streaming, have diverse and (often) stringent network requirements with respect to throughput, latency, etc. that demands a certain \gls{qos} level and availability from the communications network. 

By predicting \gls{qos} and providing in-advance notifications, \gls{pqos} emphasizes safety of road users, reliability in information exchange, end-user experience and satisfaction. Furthermore, \gls{pqos} can provide context awareness to applications depending on the driving conditions and users' needs. This can regulate application-specific survival time and ensures that the service is either continued or gracefully degraded~\cite{Boban2021}. For example, in high latency scenarios, autonomous vehicles can be proactive by either decelerating, changing lanes or handover control to the driver to avoid an accident. 

Predicting \ac{qos} is indeed a complex task, further exacerbated by the high relative vehicle speeds, vehicle density (based on time of the day) and channel conditions (based on region - rural, urban and highway) prevalent in \ac{v2x} scenarios. As part of \ac{nwdaf}, 3GPP Release 16~\cite{TS23.287} introduced an in-advance, subscribe/request-based notification system to notify \ac{v2x} applications of estimated or expected change in \ac{qos}, with the possibility to switch to an \ac{aqp} when \ac{qos} degrades. 
However, this does not ensure service continuity, as the application is notified once the change occurs.

\ac{qos} can be predicted either on the \ac{ue}-side~\cite{Figueroa2020, Moreira2020, Gutierrez2021, Skocaj2023, Vishakha2023}, or the network-side~\cite{Riihijarvi2018}, or a combined approach~\cite{Kousaridas2021}. Majority of the research for \gls{pqos} includes prediction by training models on historical data of the \gls{ue}, i.e., ego vehicle itself (henceforth, vehicle and device will be used interchangeably). However, authors in~\cite{Tomasz2021} proposed a \ac{d2d} coverage prediction framework to predict \ac{qos} by using the received signal strength measurements from the lead device. Initial findings in~\cite{Palaios2021-1} indicate that the data from the lead vehicle holds the potential to improve the \gls{pqos} of the ego vehicle. Authors in~\cite{Palaios2021-2} highlight a new set of features from the lead vehicle and network operator to improve the performance of a \gls{rf} model to predict both uplink and downlink throughput of the ego vehicle and over different prediction horizons. A similar study~\cite{Noor2024} shows improvement throughput prediction of the XGBoost regressor by selecting a different combination of features from the lead vehicle and the network operator and over different time horizons. In this paper, we intend to build upon the insights of~\cite{Palaios2021-2} and~\cite{Noor2024}, both of which have conducted their studies using a dataset collected in a highway environment.
However, unlike~\cite{Palaios2021-2}~\cite{Noor2024}, we aim to focus on improving \gls{qos} by means of instantaneous downlink throughput prediction in urban environments, i.e., focusing on the ego vehicle's current state or in the very near future, by using both current (temporally-aligned) and historical (spatially-aligned) data from the lead device. Furthermore, we train 3 different types of \gls{ml} models. In this regard, we pose the following research questions, the answers to which form the basis of our contributions:

\begin{itemize}
    \item Can the data from the lead vehicle improve the \gls{pqos} of the ego vehicle in an urban setting?
    \item Can the historical data, i.e., temporal difference between the lead and ego vehicle at the same geo-coordinate be utilized to improve the \gls{pqos} of the ego vehicle?
    \item How does feature engineering the datasets, i.e., introducing relative features impact the model performance?
    \item If the selection of features from the lead vehicles shows an improvement in model performance, is it model-agnostic?
\end{itemize}

We are using the Berlin V2X dataset~\cite{BerlinV2XDataset}, primarily, as the dataset includes vehicles driving in close proximity in an urban setting. Moreover, it is publicly available ensuring reproducibility of results. At the time of writing this paper, this dataset has been previously used to predict downlink throughput by training a \gls{lgbm} model on one operator's data and evaluating on another operator's data~\cite{Gotseva2024}; identifying nodal points in \gls{v2x} networks by predicting data rate~\cite{Saravanan2024}; and handling concept drifts when predicting \gls{qos} in resource-constrained \gls{v2x} environments~\cite{Drainakis2024}.


The rest of the paper is organized as follows. Section \ref{expl_data_analysis} provides an in-depth correlation analysis of the different features of both the ego and lead device. The correlation analysis helps select features for training the \gls{ml} models described in Section \ref{ml_model}. Section \ref{results} delivers answers to the aforementioned research questions and lastly, Section \ref{conclusion} concludes the paper and provides insights into future research directions.

\section{Exploratory Data Analysis}
\label{expl_data_analysis}

\begin{table*}[!ht]
\rowcolors{1}{gray!50}{white}
\caption{\label{tab:berlin-v2x-overview} Berlin V2X Cellular Dataset - Overview}
\begin{tabular}{|l|cccc|cccc|}
\hline
\textbf{Direction}       & \multicolumn{4}{c|}{\textbf{Downlink}}                                                                                              & \multicolumn{4}{c|}{\textbf{Uplink}}                                                                                                    \\ \hline
\textbf{Operator IDs}    & \multicolumn{2}{c|}{1}                                                 & \multicolumn{2}{c|}{2}                            & \multicolumn{2}{c|}{1}                                                   & \multicolumn{2}{c|}{2}                              \\ \hline
\textbf{Device IDs}      & \multicolumn{2}{c|}{pc1, pc4}                                          & \multicolumn{2}{c|}{pc2, pc3}                     & \multicolumn{2}{c|}{pc1, pc4}                                            & \multicolumn{2}{c|}{pc2, pc3}                       \\ \hline
\textbf{Target Datarate [kBit/s]} & \multicolumn{1}{c|}{400} & \multicolumn{1}{c|}{350,000}           & \multicolumn{1}{c|}{400} & 350,000           & \multicolumn{1}{c|}{400}   & \multicolumn{1}{c|}{75,000}            & \multicolumn{1}{c|}{400}   & 75,000            \\ \hline
\textbf{Measurement ID}  & \multicolumn{1}{c|}{3, 4}   & \multicolumn{1}{c|}{0, 1, 2, 14, 15, 16} & \multicolumn{1}{c|}{1, 2, 14}   & 0, 3, 4, 15, 16 & \multicolumn{1}{c|}{8, 9, 10} & \multicolumn{1}{c|}{5, 6, 7, 11, 12, 13} & \multicolumn{1}{c|}{8, 9, 10} & 5, 6, 7, 11, 12, 13 \\ \hline
\end{tabular}
\end{table*}

The Berlin V2X dataset captures cellular data transmission in both directions - uplink and downlink. A brief overview is provided in Table \ref{tab:berlin-v2x-overview}. For our analysis below, we focus on Measurement IDs 0 to 11 to avoid cross-operator influence (Measurement IDs 12 to 16 includes 2 devices per run, each with a different operator ID). The driving pattern ensures that for every operator, each of these measurement IDs includes a lead device and an ego device. For each device, the dataset is grouped using measurement ID to ensure  homogeneity w.r.t. device IDs, target datarate, operator ID and direction of data transmission. Furthermore, we limit our findings to the higher target datarate in downlink (termed as scenario A3D) for a single operator (ID-1), however, the same methodology can be easily extended to any combination of operator, direction and target datarate.



\subsection{Correlation - ego device} 
By analysing the correlation values between the target variable, i.e., datarate of the ego vehicle, and the features of the ego vehicle, we aim to identify the key features from the ego device to train the \gls{ml} models.

\begin{figure}[!ht]
    \centering
    \includegraphics[width=0.9\columnwidth]{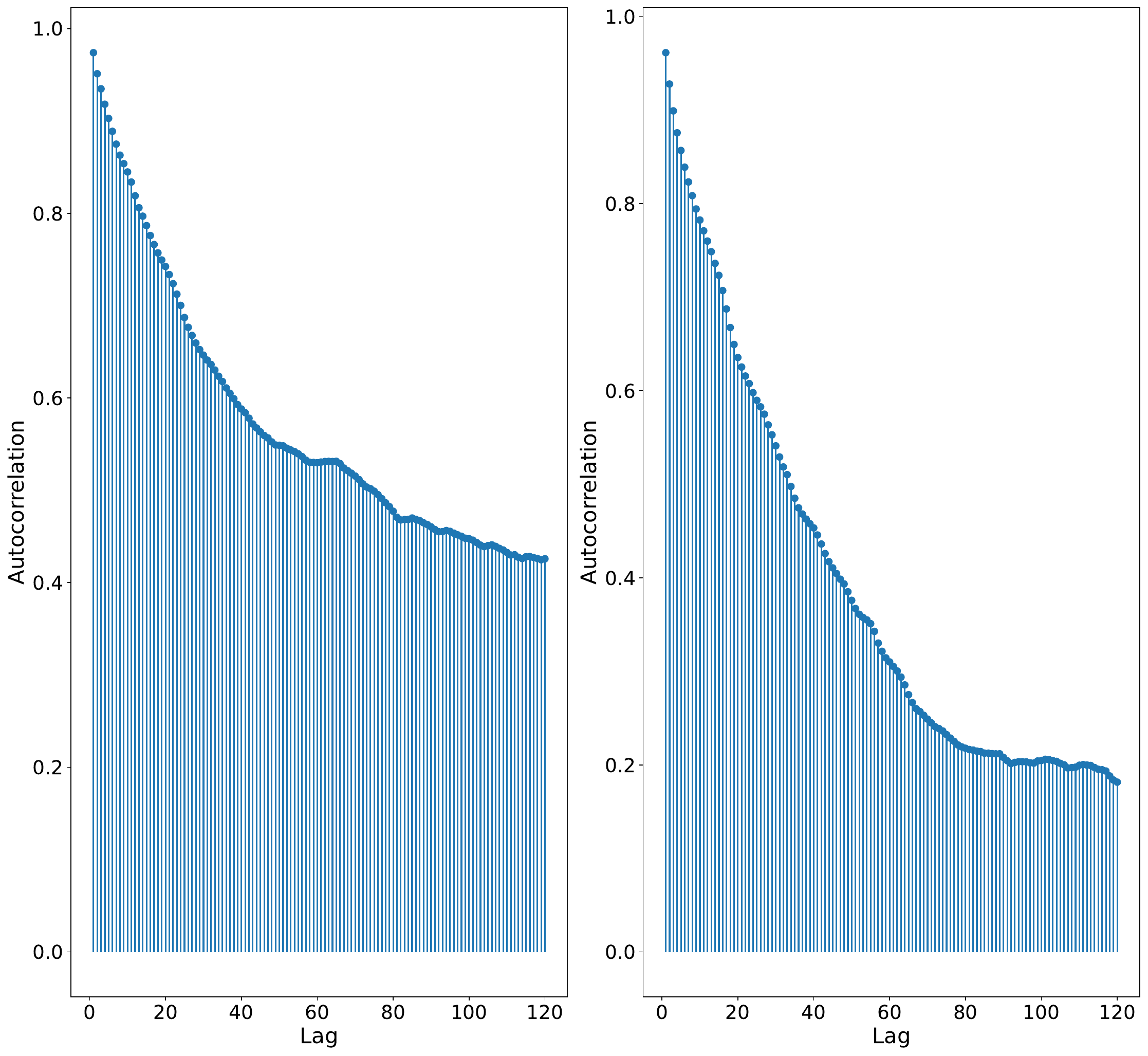}
    \caption{Autocorrelation of the datarate for the measurement IDs 0 (left) and 1 (right)}
    \label{fig:Autocorrelation}
    \end{figure}

\textbf{Autocorrelation} - Features with a strong autocorrelation are more likely to be predictable as it indicates a temporal relationship. Autocorrelation of the target variable helps identify if the lagged values of the target variable itself can be included as a predictor.
Figure~\ref{fig:Autocorrelation} show the autocorrelation for the datarate for a lag up to 120 seconds. The autocorrelation in the figures shows in both measurements a different decline rate, passing 0.5 at a lag of around 80 seconds for ID 0 and 30 seconds for ID 1. This indicates that autocorrelation is initially high, suggesting strong short-term dependencies, but then decreases, implying that the influence of past values diminishes over time.

\begin{figure}[!ht]
    \centering
    \includegraphics[width=\columnwidth]{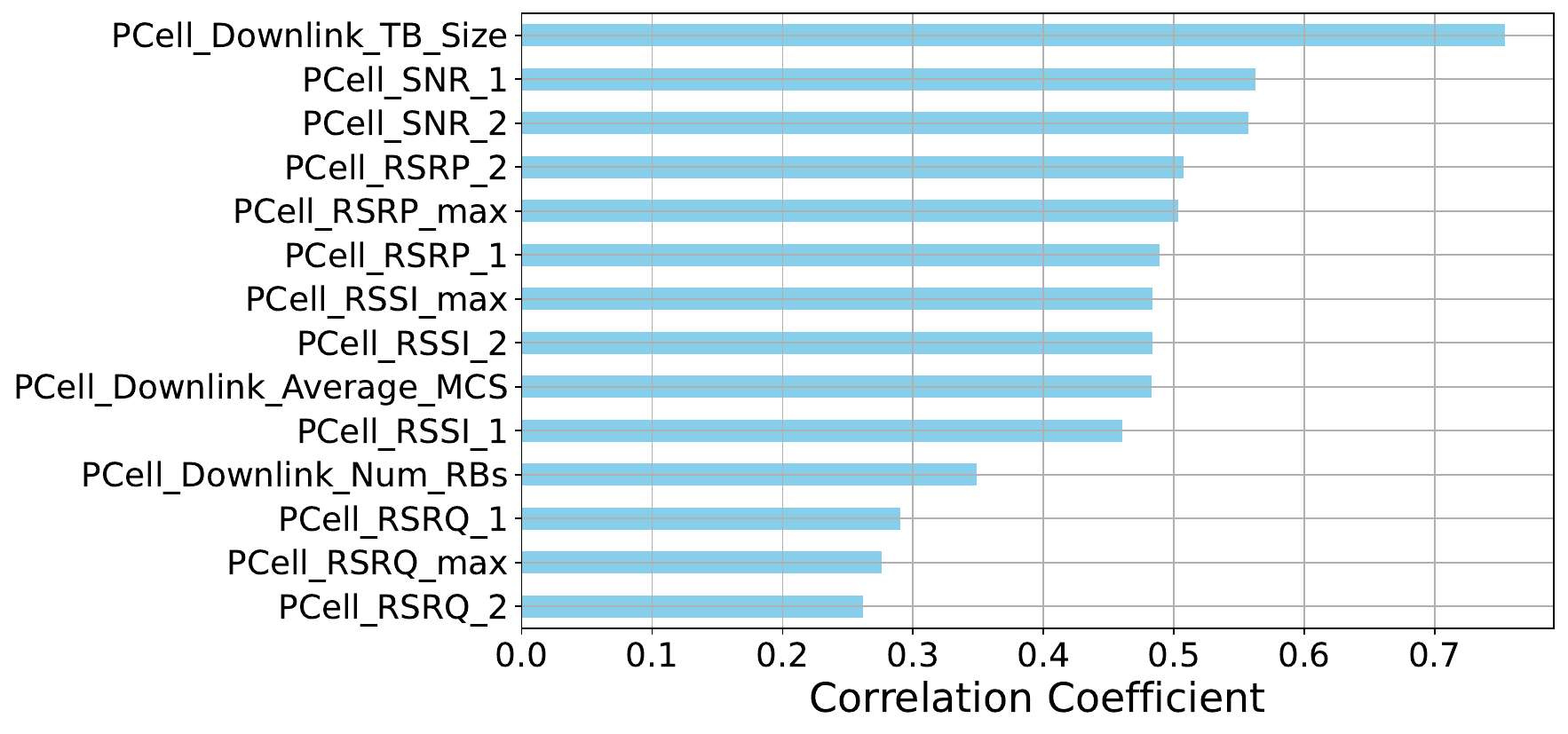}
    \caption{Features of the ego vehicle with the highest cross-correlation with the datarate.}
    \label{fig:cross_correlation_ego}
\end{figure}

\textbf{Cross-correlation} - It provides insight into the relationship between the target variable and the different features of the ego vehicle. 
Figure~\ref{fig:cross_correlation_ego} show the cross-correlation between the datarate and the most important features in the downlink for operator 1. 
The analysis indicates the transport block size (TB\_size) has the highest correlation to the datarate. Furthermore, network \glspl{kpi} such as \gls{rsrp} and \gls{snr} show relatively high correlation, making them promising input features for the prediction models. 

\subsection{Correlation - ego and lead devices}
To help identify the features of the lead device that could improve the throughput prediction for the ego device, we perform correlation analysis by aligning the devices both temporally and spatially, as follows:

\textbf{Temporally Aligned}
The devices are aligned in time in the dataset, and are directly used to evaluate the correlative measures with ever changing distances between the vehicles. 
\begin{enumerate}
    \item Same Features: 
    We aim to evaluate the features of the ego vehicle for statistical similarity against the same features of the lead vehicles (see Figure~\ref{fig:dwsamefeatCorr}). The high correlation values of the geo-coordinates clearly indicate that the vehicles travel the same route (which is the case here). Network \glspl{kpi} such as \gls{rsrp}, \gls{rsrq} and \gls{rssi} show medium to high correlation between the two devices indicating that the vehicles share more or less similar network conditions, i.e., signal fading, channel response, or congestion.  
        
    \begin{figure} 
        \centering 
        \includegraphics[width=0.8\columnwidth]{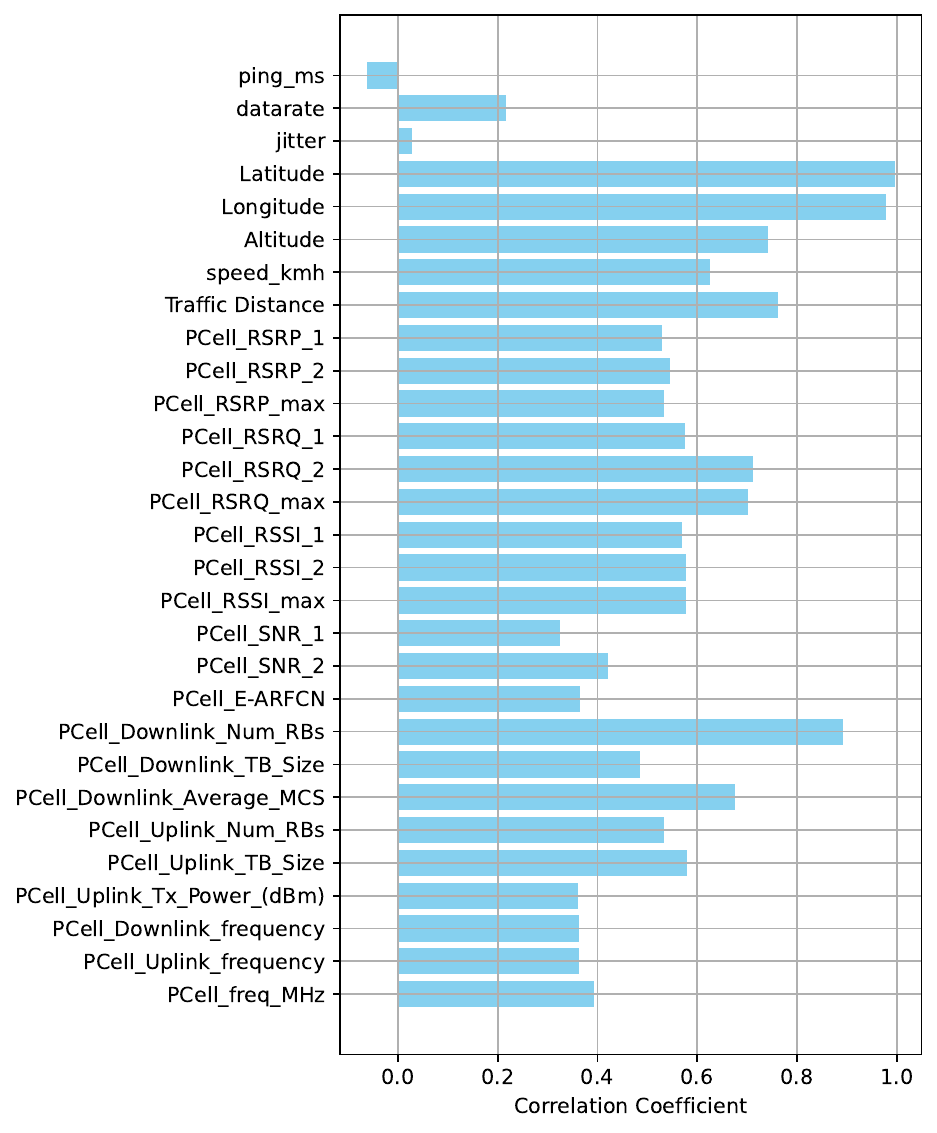}
        \caption{Correlation between the same features of the lead and ego vehicles.}
        \label{fig:downdifffeatCorr}
    \end{figure}
        
    \item Different Features: The datarate of the ego device is correlated to the features of the lead device. 
    Figure~\ref{fig:dwsamefeatCorr} illustrates these correlative patterns in the downlink direction. Here, relatively lower measures are observed for the known signal strength indicators. However, a high correlation between the lead and ego vehicle's datarate can be observed. 
    
    
\begin{figure}[ht]    
    \centering
    \includegraphics[width=0.8\columnwidth]{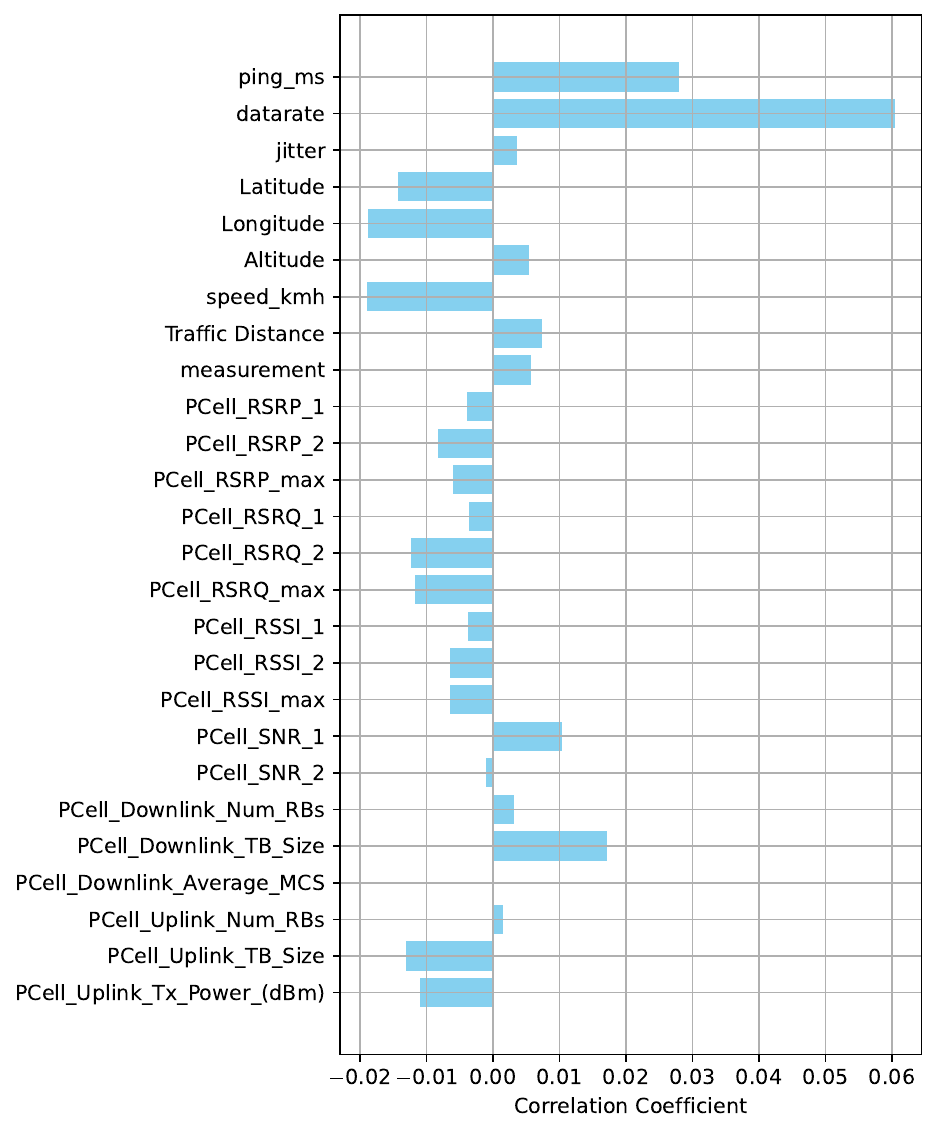}
    \caption{Correlation between the datarate of the ego vehicle and features of the lead vehicle.}
    \label{fig:dwsamefeatCorr}
\end{figure}
\end{enumerate}

\textbf{Spatially Aligned}
Devices are spatially aligned to understand the correlation between current values of features of the ego device and historical values of features of lead device at (approximately) the same geo-coordinate. Practically, this can provide an crucial insight into the relevancy of lead vehicle's historical data to be stored. To align the devices spatially, the geodesic distance between the devices is computed. For each geo-coordinate of the ego device, the closest location pairs - \textit{\{ego, lead\}} are extracted that fall within a distance threshold of 0-20 m between the devices. This operation is performed for increasing time offsets - in minute increments (i.e., $0-1, 1-2, \cdots$). For each time offset range the data of the lead device is filtered, of which only the geo-coordinates within the threshold distance to the ego device are considered.
\begin{enumerate}
    \item Same Features -  An equal-weighted average of the pair-wise feature correlation (between lead and ego devices) across all measurement IDs is shown in Figure~\ref{fig:op1_spat_same_features_downlink} .
    This is achieved by computing correlation for every measurement ID individually and then averaging to get the weighted average. The pair-wise features, especially the signal strength indicators, exhibit medium to generally, high correlation values indicating that the environment impacts both devices similarly, i.e., signal strength fluctuations due to obstacles. This is true for both time offset ranges, i.e., 0-1 and 1-2 mins, with correlation increasing as more historical data from the lead vehicle becomes available, i.e., data with a larger temporal gap with the ego vehicle. This trend cannot be mapped any further due to lack of sufficient data points for larger time offset ranges, i.e., 2-3, 3-4 mins, and so on.
    
    \begin{figure}[!ht]
        \centering
        \includegraphics[width=0.8\columnwidth]{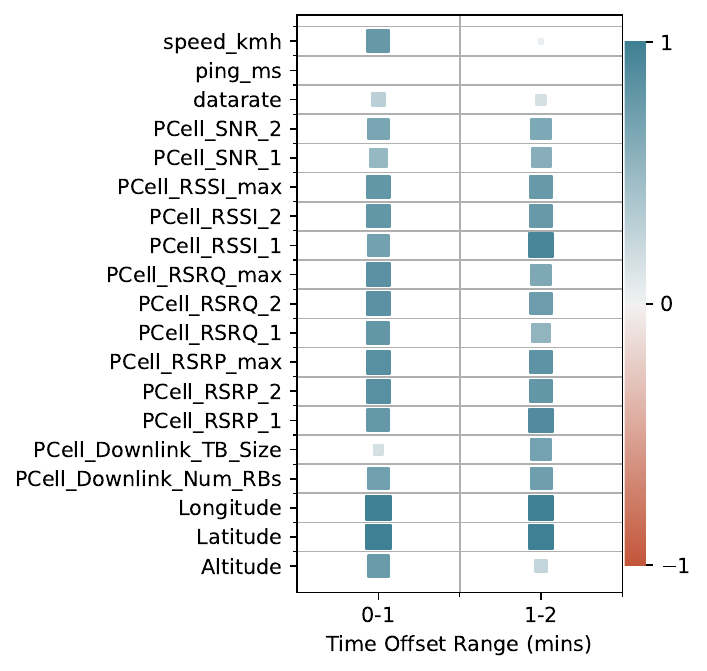}
        \caption{Pair-wise correlation values for spatially aligned ego and lead vehicles. The colour bar depicts the correlation coefficient.}
        \label{fig:op1_spat_same_features_downlink}
    \end{figure}
    
    \item Different Features - Correlation analysis between datarate of the ego vehicle and all features of the lead vehicle shows no features with medium to high correlation, apart from geo-coordinates, traffic conditions and weather data.  
\end{enumerate}

\section{Machine Learning Models}
\label{ml_model}

This section provides a concise overview of the dataset preparation and the criteria for model selection. Based on the correlation analysis, we devise the following datasets: 
\begin{itemize}
    \item \textbf{EGF} - only features from the ego device (baseline)
    \item \textbf{EGLT} - features from EGF and features from the lead device when temporally aligned
    \item \textbf{EGLS} - features from EGF and features from the lead device when spatially aligned 
\end{itemize}
 For these datasets, we define the selected features and the \gls{ml} models used below.

\subsection{Feature Engineering}
The in-depth correlation analysis yields a multitude of features of interest for each dataset. Features relating to weather and traffic conditions are dropped, because they have a lower sampling rate. Even though certain features of the lead vehicle do not show direct correlation with the datarate of the ego vehicle, these features have high correlations with their respective pairs (i.e., same feature from ego vehicle). 

To avoid model overfitting, similar and highly collinear features are removed, for example, we retain \textit{PCell\_RSRP\_max} and drop its highly correlating (> 0.7) counterparts \textit{PCell\_RSRP\_1} and \textit{PCell\_RSRP\_2}. Furthermore, to provide spatio-temporal context to the models, we devise $delta\_t$, $delta\_s$ and $delta\_v$ which represent the difference in time, distance and speed between the ego and the lead vehicles. Due to the introduction of $delta\_s$ the geo-coordinates were dropped from the features. Note that $delta\_t$ is not used for the EGLT scenario as the two vehicles are temporally aligned. Table \ref{tab:features_of_interest_ml} shows the features selected for each scenario (feature names are taken directly from the dataset).

To help the models understand the dynamics between the two vehicles, we also transform features and provide them as alternate variants to the scenarios EGLT and EGLS:
\begin{itemize}
    \item Difference Features - the difference captures the relative dynamics between the ego and lead vehicles by combining the same features from the 2 vehicles. For example, $PCell\_RSRP\_max$ is computed as the difference between $lead\_PCell\_RSRP\_max$ and $ego\_PCell\_RSRP\_max$.
    \item Ratio Features - this captures the proportional relationship between the features of the 2 vehicles. For example, $PCell\_RSRP\_max$ is computed as the ratio between $lead\_PCell\_RSRP\_max$ and $ego\_PCell\_RSRP\_max$.
\end{itemize}
By applying difference features to EGLT and ratio features to EGLS, the goal is to gather insights on the potential impact of each feature engineering on model performance in different contexts, rather than conducting a comprehensive analysis by testing their generalizability across all datasets. Henceforth, the transformed datasets are referenced with a suffix - EGLT-Diff and EGLS-Ratio indicating the feature transformation used.

\begin{table*}[!ht]
\rowcolors{1}{gray!50}{white}
\caption{\label{tab:features_of_interest_ml} Input features and size for datasets used to train \gls{ml} models}
\centering
\resizebox{\textwidth}{!}{%
\begin{tabular}{|c|c|l|c|}
\hline
    \textbf{Scenario} & \textbf{Vehicle} & \textbf{Features} & \textbf{Dataset Size}
    \\ \hline
\textbf{EGF}  & 
    \begin{tabular}[c]{@{}l@{}} 
    Ego only
    \end{tabular} & 
    \begin{tabular}[c]{@{}l@{}}
    PCell\_SNR\_1, PCell\_SNR\_2, PCell\_RSRP\_max,\\ PCell\_RSRQ\_max, PCell\_RSSI\_max, PCell\_Downlink\_TB\_Size, \\ 
    PCell\_Downlink\_Num\_RBs, PCell\_Downlink\_Average\_MCS
    \end{tabular} &
    \begin{tabular}[c]{@{}l@{}} 
    9699
    \end{tabular}
\\ \hline
\textbf{EGLT} &
    \begin{tabular}[c]{@{}l@{}}
    Ego and Lead
    \end{tabular} &
    \begin{tabular}[c]{@{}l@{}}
    lead\_datarate, PCell\_Downlink\_TB\_Size, PCell\_Downlink\_Tx\_Power,\\  PCell\_Downlink\_Num\_RBs, PCell\_Downlink\_Average\_MCS, \\ PCell\_RSSI\_max, delta\_v, delta\_s 
    \end{tabular} &
    \begin{tabular}[c]{@{}l@{}} 
    9442
    \end{tabular}
\\ \hline
\textbf{EGLS} &
    \begin{tabular}[c]{@{}l@{}}
    Ego and Lead
    \end{tabular} &
    \begin{tabular}[c]{@{}l@{}}
    PCell\_SNR\_1, PCell\_SNR\_2, PCell\_RSRP\_max, PCell\_RSRQ\_max, \\ PCell\_RSSI\_max, PCell\_Downlink\_Num\_RBs,\\ PCell\_Downlink\_Average\_MCS, 
    delta\_v, delta\_s, delta\_t
    \end{tabular} &
    \begin{tabular}[c]{@{}l@{}} 
    6146
    \end{tabular}
\\ \hline
\end{tabular}%
}
\end{table*}

\subsection{Dataset Preparation: Scaling and Splitting}
After selecting, reducing and transforming the features, we normalize the datasets using min-max scaling to a range of $0$ to $1$. This step is necessary because the features have varying units and scales. Min-max scaling ensures that all features contribute proportionally and prevents features with larger numerical ranges from dominating the learning process. The datasets are then split into an 80:20 ratio for training and testing purposes. To preserve the temporal structure of the data, no shuffling is performed during the split.

\subsection{\gls{ml} Models}
The choice of \gls{ml} models depends on a number of factors, including but not limited to the nature and size of available data, complexity of relationships between the features, task at hand (prediction, clustering, etc.). In this study, we aim to predict the downlink datarate of the ego vehicle, which is a multivariate time-series regression task. Thus, we select 3 types of \gls{ml} models suited for this task - tree-based XGBoost, feedforward-based \gls{cnn}, and recurrent-based \gls{lstm}. 

Data is reordered for both \acs{cnn} and \acs{lstm} models with a lookback time window of \si{60} timesteps, i.e., \si{60} timesteps are used to predict the next value and are trained for \si{100} epochs.



\section{Results}
\label{results}

\begin{figure}[!ht]
    \centering
    \includegraphics[width=\columnwidth]{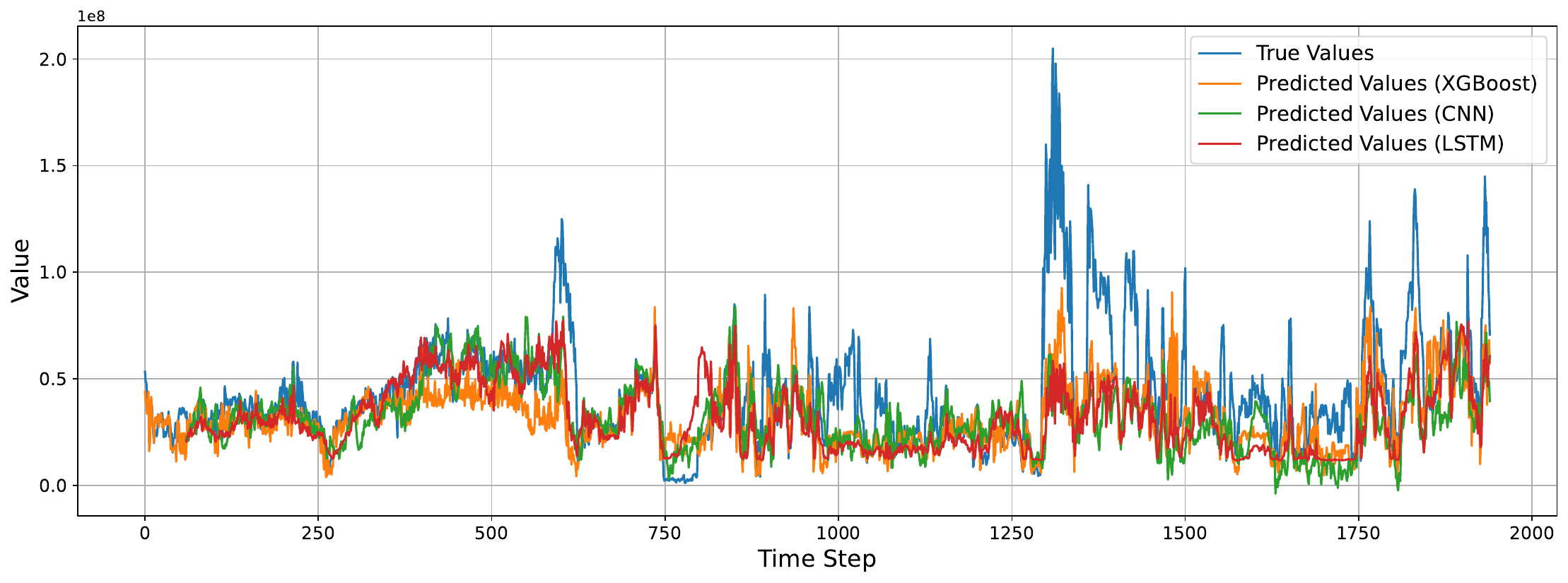}
    \caption{True and predicted datarate over time for the test set for all 3 models on the EGF dataset}
    \label{fig:EGF}
    \end{figure}

\begin{figure}[!ht]
    \centering
    \includegraphics[width=\columnwidth]{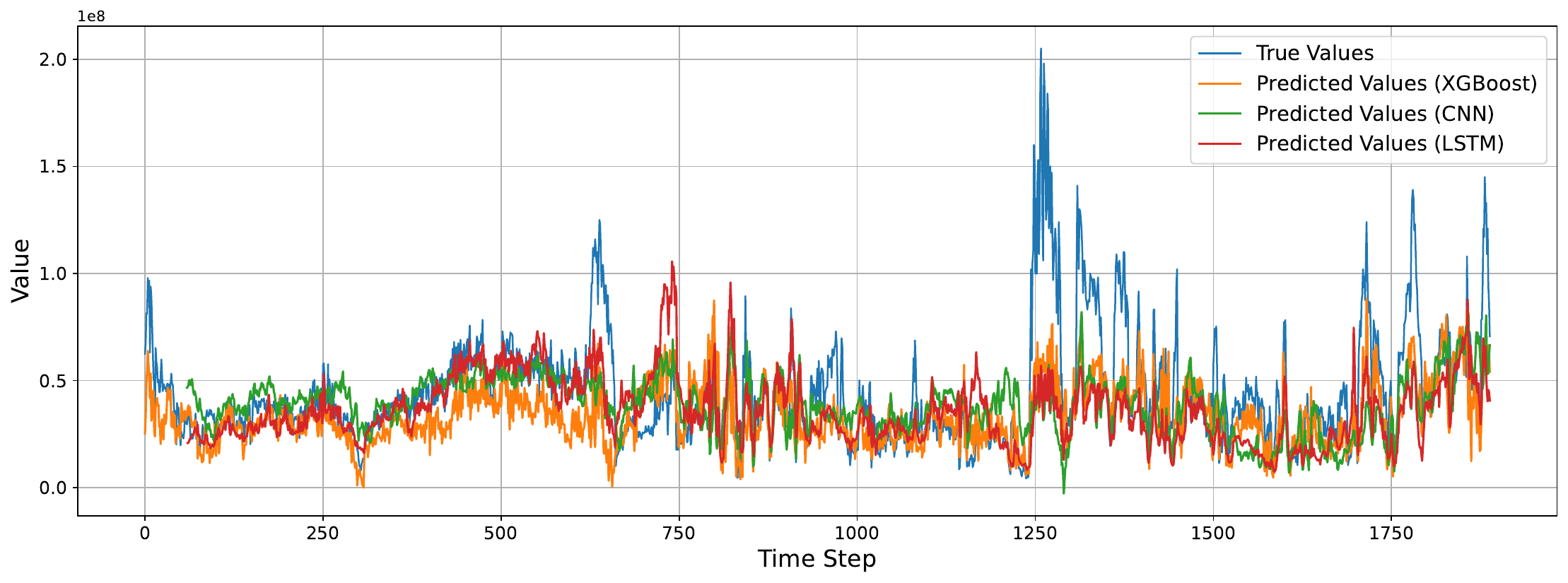}
    \caption{True and predicted datarate over time for the test set for all 3 models on the EGLT dataset}
    \label{fig:EGLT}
    \end{figure}

\begin{figure}[!ht]
    \centering
    \includegraphics[width=\columnwidth]{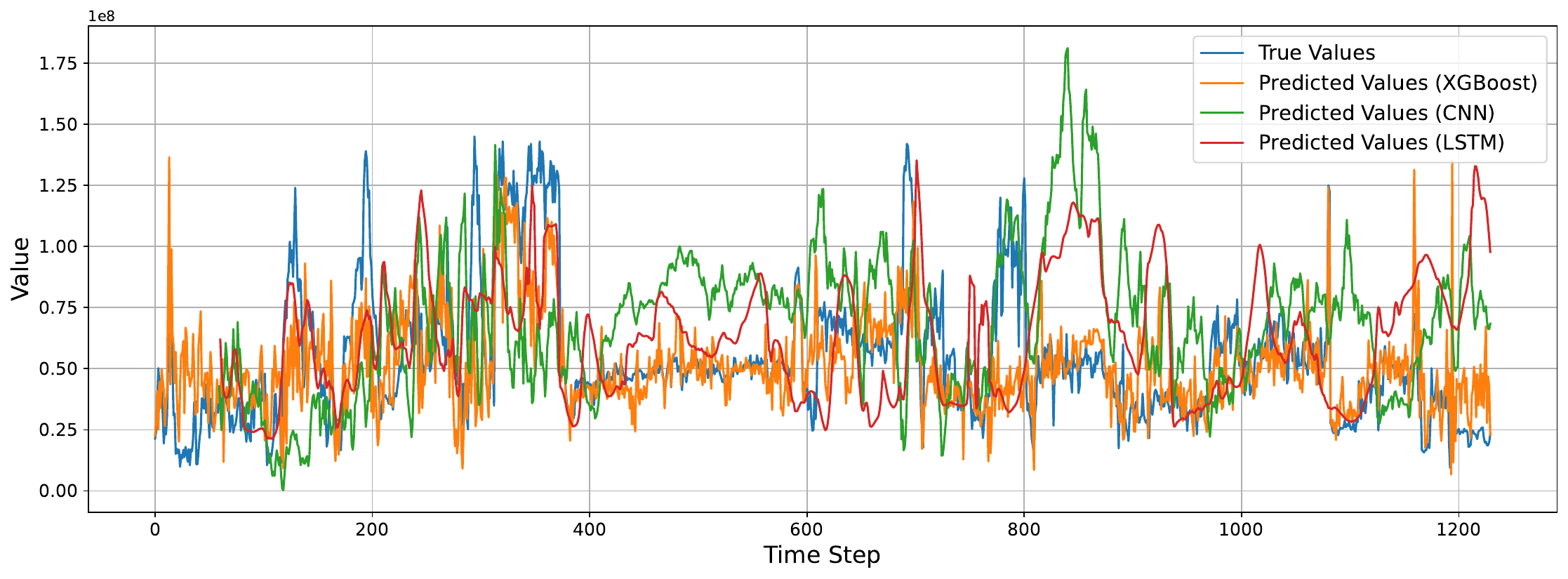}
    \caption{True and predicted datarate over time for the test set for all 3 models on the EGLS-Ratio dataset}
    \label{fig:EGLS_Ratio}
    \end{figure}

The models are trained and tested on all datasets for 50 runs to reduce variability and ensure reliable performance assessment of the models. One such run for the datasets EGF, EGLT and EGLS-Ratio is seen in Figure~\ref{fig:EGF}, Figure~\ref{fig:EGLT} and Figure~\ref{fig:EGLS_Ratio}, respectively. These figures yield a comparative visual inspection of model predictions for the datasets in question. For instance, it is evident in Figures~\ref{fig:EGF} and~\ref{fig:EGLT}, all models have similar prediction patterns when compared to the actual values. However, Figure~\ref{fig:EGLS_Ratio} shows that \gls{lstm} and \gls{cnn} fail to capture the underlying relationships in the EGLS-Ratio dataset. This can be attributed to the nature of the dataset transformation, i.e., ratio, which reduces the variability in the data. This decline in model performance is also noted in the model performance metrics listed in Table~\ref{tab:model_metrics}.

To compare the models, we consider \gls{mae}, \gls{smape} and \gls{rmse} as the error metrics. We use a  modified version of \gls{smape}, by introducing a small $\epsilon$ of $e^{-8}$, so as to mitigate the near zero values generated in the denominator by min-max scaling.
\begin{equation*}
    SMAPE = \frac{1}{n} \sum_{t=1}^{n}\frac{|F_t - A_t|}{\max(\epsilon, (|{A_t}| + |{F_t}|)/2)}
\end{equation*}
where $A_t$ is the actual value, $F_t$ is the predicted value and $n$ is the total number of actual values. 

EGF is considered as the baseline in this study with only features from the ego vehicle itself. It serves as a reference point, i.e., reference error metrics for all models. When comparing other datasets to EGF, it is evident that all error metrics are lower, i.e., with the introduction of features from the lead vehicle leads to improved model performance. Specifically, EGLS performs better for all models indicating the importance of storing spatially-aligned historical data. For instance, the biggest improvement in \gls{mae} (37.87\%), \gls{smape} (54.24\%) and \gls{rmse} (37.15\%) can be seen with the XGBoost model in the EGLS dataset. The transformed dataset EGLT-Diff consistently yields lower error metrics, even lower than the original EGLT dataset. This indicates by introducing the difference transformation helped simplify the dataset, and made it easier for the models to capture spatio-temporal aspects of the vehicles. 

When comparing the models, XGBoost outperforms \gls{cnn} and \gls{lstm} on most metrics and datasets. \gls{lstm} shows comparable performance to XGBoost in some datasets, but generally, has higher error metrics. Furthermore, \gls{lstm} shows lower \gls{smape} values than \gls{cnn} in some datasets as it can better leverage the sequential nature of the datasets than \gls{cnn}. Even though \gls{cnn} struggles to perform in comparison to the other models, it still offers consistent improvement when compared to the EGF dataset, hence, confirming that the improvement in performance is model-agnostic.






    
\begin{table*}[!ht]
\centering
\caption{ML Model Metrics}
\label{tab:model_metrics}
\resizebox{2\columnwidth}{!}{%
\begin{tabular}{|l|ccc|ccc|ccc|}
\hline
\multirow{2}{*}{\textbf{Dataset}} & \multicolumn{3}{c|}{\textbf{XGBoost}}                                                                                                                            & \multicolumn{3}{c|}{\textbf{CNN}}                                                                                                                                & \multicolumn{3}{c|}{\textbf{LSTM}}                                                                                                                               \\ \cline{2-10} 
                                  & \multicolumn{1}{l|}{\textbf{MAE}} & \multicolumn{1}{l|}{\textbf{SMAPE}} & \textbf{RMSE} & \multicolumn{1}{l|}{\textbf{MAE}} & \multicolumn{1}{l|}{\textbf{SMAPE}} & \textbf{RMSE} & \multicolumn{1}{l|}{\textbf{MAE}} & \multicolumn{1}{l|}{\textbf{SMAPE}} & \textbf{RMSE} \\ \hline
\textbf{EGF}  & 
\multicolumn{1}{l|}{0.0705} & \multicolumn{1}{l|}{0.4096} & 0.1074 & 
\multicolumn{1}{l|}{0.0938} & \multicolumn{1}{l|}{0.7092} & 0.1378 &
\multicolumn{1}{l|}{0.0784} & \multicolumn{1}{l|}{0.6243} & 0.1199 \\ 
\hline
\textbf{EGLT} & 
\multicolumn{1}{l|}{0.0721} & \multicolumn{1}{l|}{0.3778} & 0.1090 & 
\multicolumn{1}{l|}{0.0883} & \multicolumn{1}{l|}{0.5996} & 0.1277 & 
\multicolumn{1}{l|}{0.0772} & \multicolumn{1}{l|}{0.5728} & 0.1172 \\
\hline
\textbf{EGLT-Diff} & 
\multicolumn{1}{l|}{0.0596} & \multicolumn{1}{l|}{0.3284} & 0.0756 & 
\multicolumn{1}{l|}{0.0816} & \multicolumn{1}{l|}{0.5541} & 0.1064 & 
\multicolumn{1}{l|}{0.0754} & \multicolumn{1}{l|}{0.5154} & 0.0977 \\
\hline
\textbf{EGLS} & 
\multicolumn{1}{l|}{0.0438} & \multicolumn{1}{l|}{0.1874} & 0.0675 & 
\multicolumn{1}{l|}{0.0794} & \multicolumn{1}{l|}{0.4849} & 0.1109 & 
\multicolumn{1}{l|}{0.0704} & \multicolumn{1}{l|}{0.4797} & 0.1002 \\ 
\hline
\textbf{EGLS-Ratio} & 
\multicolumn{1}{l|}{0.0711} & \multicolumn{1}{l|}{0.2963} & 0.1043 & 
\multicolumn{1}{l|}{0.1568} & \multicolumn{1}{l|}{0.5707} & 0.1997 & 
\multicolumn{1}{l|}{0.1213} & \multicolumn{1}{l|}{0.4848} & 0.1631 \\
\hline
\end{tabular}%
}
\end{table*}

\section{Conclusion and Future Work}
\label{conclusion}
This paper predicts the downlink throughput for the ego device and aims to understand if the model performance can be improved by adding features from the lead device. To this end, we employ the Berlin V2X dataset. The dataset is filtered into 3 primary datasets - EGF representing features from the ego device only, EGLT representing features from ego and lead devices when temporally aligned and EGLS representing features from ego and lead devices when spatially aligned. Besides this, we transform the features of the EGLT and EGLS datasets using differences and ratios, respectively, to help capture the underlying spatio-temporal relationships between the 2 devices. By employing different types of \gls{ml} models, namely, XGBoost, \gls{cnn} and \gls{lstm}, we show that the model performance is improved by addition of features from the lead device across all datasets (barring the EGLS-Ratio dataset). Moreover, the improvement is model-agnostic, with the biggest improvement shown in the EGLS dataset by the XGBoost model boasting error values of \gls{mae} 0.0438 (37.87\%), \gls{smape} 0.1874 (54.24\%) and \gls{rmse} 0.0675 (37.15\%). As part of the future work, we intend to improve the dataset by generating synthetic data conforming with the available real-world measurements. We would also fine tune the models based on the time of the day and the area in which the devices are present. Furthermore, we would investigate the throughput prediction and improvement when multiple mobile network operators are present.  

For successful real-world deployment, certain practical challenges must be addressed. For example, sharing of real-time information raises privacy concerns and questions regarding dissemination of data, accessibility to this data and the frequency of transmission should be addressed. Moreover, deployment of ML algorithms in a dynamic vehicular environment raises its own set of challenges, including but not limited to model adaptability, computational efficiency and integration with existing vehicular networks and systems. However, promising solutions have been explored, like protocol designs specific for short range V2V communication. Federated learning techniques have also been proposed to close the data privacy gap. 5G additionally provides resource abstractions like dedicated network slices for the necessary low latency connectivity. This connectivity will be augmented with other non-terrestrial formations and road side units, such that workload can be offloaded when the need for seamless deployment of ML algorithms arises.



\section*{Acknowledgement}
This work has been supported by the Federal Ministry of Education and Research of the Federal Republic of Germany as part of the Open6GHub project (16KISK004, 16KISK003K) and the SUSTAINET\_guarDian (16KIS2239K). The authors alone are responsible for the content of the paper.

\bibliographystyle{IEEEtran}


\end{document}